# Extension of Three-Variable Counterfactual Casual Graphic Model：from Two-Value to Three-Value Random Variable

Jingwei Liu [1]
School of Mathematics and System Sciences,
Beihang University,
Beijing, P.R China, 100191.

## Abstract

The extension of counterfactual causal graphic model with three variables of vertex set in directed acyclic graph (DAG) is discussed in this paper by extending two- value distribution to three-value distribution of the variables involved in DAG. Using the conditional independence as ancillary information, 6 kinds of extension counterfactual causal graphic models with some variables are extended from two–value distribution to three–value distribution and the sufficient conditions of identifiability are derived.

## 1. Introduction

Causal graphic model is one of the most important model based on DAG in many research fields, such as biological medicine , social science, epidemiology, machine learning and inference, software reliability [2], and so on [1-14]. The research significance of causal model lies in the causal relationship between the research objects, rather than just statistical correlation, thus to infer the behavior or strategy effects on the study object. Two of the important DAG models, counterfactual model proposed by Rubin(1978) [1] and causal diagram model proposed by Pearl [2,3], outline a framework for causality inference analysis.

The causal effect of counterfactual model is obtained by "intervention" of control variables, to analyze the identification of the causality. If causal effect is identifiable, it is possible to directly calculate the causal effect from the observational data, without redesigning the experiment to obtain observation data. Since the counterfactual model without any assumption or constrained conditions is not identifiable, adding the conditional independent constraints to find the identifiability of causal effects become one of the most important research problems in counterfactual causal model. Zheng, *et al.* [4,13] investigated the identification conditions of a three-variable DAG with two-value discrete distributions, in which control variable and the covariate variable is independent, and, proposed that with specific replaceability condition, the causal effect is identifiable. Liang, *et al.*[5]

---

[1]Corresponding author. jwliu@buaa.edu.cn (J.W Liu)
[2]The causal graphic models are categorized into two classes: the famous fault tree analysis(FTA) model is a DAG [9,10] , while the general dynamic causal graphic model is directed cyclic graph (DCG) [11,12] .

discussed another counterfactual causal model, in which the control variable acts on the covariate variable, and pointed out that whether the control variable takes effect on the covariate variable has no effect on the identifiability of causal effects and the assumption of replaceability. And, the sufficient and necessary conditions are given for causal effect identifiability.

However, the two kinds of counterfactual causal models discussed in [4,5,13,14] are the most simplest DAG models, whose variables are three-variable with two-value distribution. In the real world study, the value of each random variable in the causal model may have multiple values, not definitely just two values. For example, in the case of “ smoking causes lung cancer”, the deep research of the observed object, focus on not only the effect of “ Smoking or non-smoking” on “ having or not having lung cancer”, but also the degree of impacts of smoking (“ non-smoking”, “ light smoking ”, “ Heavy smoking”) on the degree of lung cancer (for example: {“ having lung cancer”, “ not having lung cancer” } or {“ not having lung cancer”, “ early stage of lung cancer”, “last stage lung cancer” }). Another example is from software reliability, the different workings status of a engine component have different impact on the other parts which have causal relationship with it, it is also necessary to extend the two-value causal model to the case of multi-value [12]. Furthermore, the measure of the operating status of certain variables tend to adopt a multi-level standard, for example, “ Excellent”, “ good”, “ medium”, “ poor” and other standards.

In this paper, we will extend the two-value three-variable DAG to three-value three-variable DAG, give the intervention definition of multi-value distribution, and derive the identification condition of 6 specific DAG models.

The rest of the paper is organized as follows: In section 2, we introduce the main notation and definitions. In section 3, Extension of three kinds of three-variable counterfactual models are discussed, where input variable is independent to instrumental variable. In section 4, Extension of three kinds of three-variable counterfactual models are discussed, where input variable is dependent to instrumental variable. Section 5 concludes the paper.

# 2. Preliminary Notation and definitions

## 2.1. DAG, Causal model and Counterfactual model

**Definition 1.**[4] DAG $G=<V,E>$ is a 2 elements set, where $V=\{V_1,V_2,\cdots,V_n\}$ is the vertices set of graph $G$, $E=\{(V_1,V_2)\,|\,V_1,V_2\in V\}$ is the set of directed edges, $(V_1,V_2)$ represents the directed edge of $V_1\rightarrow V_2$. For any vertex $V_i$ of $G$, if there is no way to start at $V_i$ and follow a sequence of edges that eventually loops back to $V_i$ again, $G$ is called a DAG (directed acyclic graph). Let $V$ represent the random variable, directed edges represent the causal relationship of variables, the DAG is called a causal graph.

**Definition 2.**[2,4] Suppose $G=<V,E>$ is a DAG. For any vertex $V_i$ of $G$, denote $pa(V_i)=\{v\,|\,v\in V,(v,V_i)\in E\}$ as the set of vertices directly point to vertex $V_i$.

$pa(V_i)$ is called the parent set of $V_i$, and $V_i$ is called the son of vertices in $pa(V_i)$. If the join distribution function $P$ of $V=(V_1,V_2,\cdots,V_n)$ satisfy,

$$P\{V_1,V_2,\cdots,V_n\}=\prod_{i=1}^{n}P\{V_i\mid pa(V_i)\}, \tag{2.1}$$

where $P\{V_i\mid pa(V_i)\}$ is the common conditional distribution. $P$ is called Markov compatibility to DAG $G$.

The motivation of this paper is to investigate the DAG $G$ of three vertices $\{X,Y,Z\}$ with multi–value discrete random variables which sample spaces are $X$ $=\{0,1,\cdots,K-1\}$, $Y$ $=\{0,1,\cdots,M-1\}$, $Z=\{0,1,\cdots,N-1\}$ respectively, where $X$ is the control variable, $Z$ is the ancillary (or instrumental) variable , $Y$ is the outcome variable. Let“$\perp$” represent “independence”. While $K=M=N=2$, it is the investigated case of [4,5,13]. And, when $K=M=2$, it is the case of [14].

Suppose $\forall v\in X\times Y\times Z, P(v)>0$. Denote all the probability measure set which is compatible to DAG $G$ as

$$\begin{aligned}P\triangleq\{P: P &\text{ is the probability measure Markov compatible}\\ &\text{to DAG } G \text{ on } X\times Y\times Z,\quad \text{and } \forall v\in X\times Y\times Z, P(v)>0.\}\end{aligned} \tag{2.2}$$

**Definition 3.** [2,4] Suppose $X$ is the control variable on the sample space $X=\{0,1,\cdots,K-1\}$ of DAG $G$. Intervention on $X$ is, for any $k$ $(1\le k\le K)$ of $i_1,\cdots,i_k(\in X)$, to force the $k\in X\setminus\{i_1,\cdots,i_k\}$ to be not happened, while the events of $X\in\{i_1,\cdots,i_k\}$ to be happened. Hence, reconstruct a counterfactual sample space or counterfactual probability measure $P_{i_1,\cdots,i_k}$. We call it the intervention distribution, and the intervention distribution is still compatible to DAG $G$.

As $X$ is a $K$–value discrete random variable, the intervention on it could lead to $C_K^1+C_K^2+\cdots+C_K^K=2^K-1$ possible combination of probability measures[3]. We only discuss the intervention on a single element value of a $K$–value random variable, thus we only discuss the causal distribution of $Y$ with the intervention on “X=0”, denote it $P_0(Y)$. When $K=2$, it is the case of intervention discussed in [4,5,13,14].

**Definition 4.** [4] For $\forall P\in P$, if the parameter in $P\to P_{i_1,\cdots,i_k}(Y)$ in Definition 3 is identifiable, the causal effect is called identifiability.

The identifiability of $P_0(Y)$ in [4,5,13,14] is the special case of $P_{i_1,\cdots,i_k}(Y)$ with $K=2$.

**Theorem 1.** [4] Generally, the causal effect with intervention on "X=0" is non-identifiable.

---

[3]If $K=1$, intervention on it is also possible in theory. However, the discussion could not be solved under the framework of [4,5,13,14]. If we can perform intervention on the 1–value distribution, we can do it on the $K$–value discrete distribution. In this paper, we only discuss the case of $K\ge 2$.

## 2.2. Review of identifiability on three-variable DAG with 2-value

To facilitate the description of our new counterfactual model, we first outline the present results on two kinds of three-variable DAG with 2-value.

The first DAG $G$ is that $X$ is control variable, $Y$ is the outcome variable, and $Z$ is the instrumental variable, their sample spaces are $X = Y = Z = \{0,1\}$. And, DAG $G$ satisfies the assumption "$X \perp Y$", there is aconclusion as follows,

**Theorem 2.** [4,13] Suppose $P_0$ satisfy one of the following condition:

(a) $X \perp Y$ ;

OR (b) $X \perp Y \mid Z$ ;

OR (c) $X \perp Y \mid Z = 0 \bigcap Y \perp Z \mid X = 1$;

OR (d) $X \perp Y \mid Z = 1 \bigcap Y \perp Z \mid X = 1$.

$P_0(y)$ is identifiable.

The second DAG $G$ is that $X$ is control variable, $Y$ is the outcome variable, and $Z$ is the instrumental variable, their sample spaces are $X = Y = Z = \{0,1\}$. And, DAG $G$ satisfies the assumption "$X \not\perp Y$", there are two conclusions as follows,

**Theorem 3.** [5] The sufficient and necessary condition of identifiability of $P_0(y)$ is $P_0$ satisfies one of the following conditions:

(a) $X \perp Y$;

OR (b) $X \perp Y \mid Z = 0 \bigcap Y \perp Z \mid X = 1$;

OR (c) $X \perp Y \mid Z = 1 \bigcap Y \perp Z \mid X = 1$.

$P_0(y)$ can be calculated from the observed distribution values of $P$.

**Theorem 4.** [5] Suppose one of the replaceability assumption of $P_0$ holds:

(a) $X \perp Y$;

OR (b) $X \perp Y \mid Z = 0 \bigcap Y \perp Z \mid X = 1$ ;

OR (c) $X \perp Y \mid Z = 1 \bigcap Y \perp Z \mid X = 1$ .

The sufficient and necessary condition $P_0(y)$ can be expressed thoroughly from the observed distribution is the observed distribution $P_0$ satisfies $(Y \perp Z \mid X = 0)_{P_0}$.

The goal of this paper is to discuss the identifiability problem after extending 2-value to 3-value counterfactual causal graphic model. Extending 3-variable with 2-value DAG to 3-variable with 3-value DAG, there are totally ($C_3^1 + C_3^2 + C_3^3 = 7$) cases, we only give the identifiability conclusions under the conditions of "$X \perp Z$" and "$X \not\perp Z$", and "$K = 3, M = 3, N = 3$", "$K = 3, M = 2, N = 3$", "$K = 2, M = 3, N = 2$" respectively, totally 6 cases of counterfactual causal graphic model. Theoretically, we can derive the identifiability conditions of $P_0(Y = k), \quad k = 2, \cdots, M - 1$, and determine the identifiability of $P_0(Y)$ and $P_0(Y = k \mid X = 1), \quad k = 1, \cdots, M - 1$. Limited to the paper length, we only discuss the identifiability of $P_0(Y = 1)$.

# 3. Extension of DAG with X,Y,Z, and " $X \perp Z$ "

## 3.1. " $X \perp Z$ ", Counterfactual causal DAG with 3-value X, 3-value Y, 3-value Z

### 3.1.1. DAG model description and identifiable instrumental information

Suppose $X, Y, Z$ are 3-value random variables, such that the sample spaces of $X$, $Y$ and $Z$ are $X = Y = Z = \{0,1,2\}$, where $X$ is control variable, $Y$ is outcome variable, $Z$ is instrumental variable. Both $X$ and $Z$ havecausal effects on $Y$, and $X \perp Z$ (Fig.1).

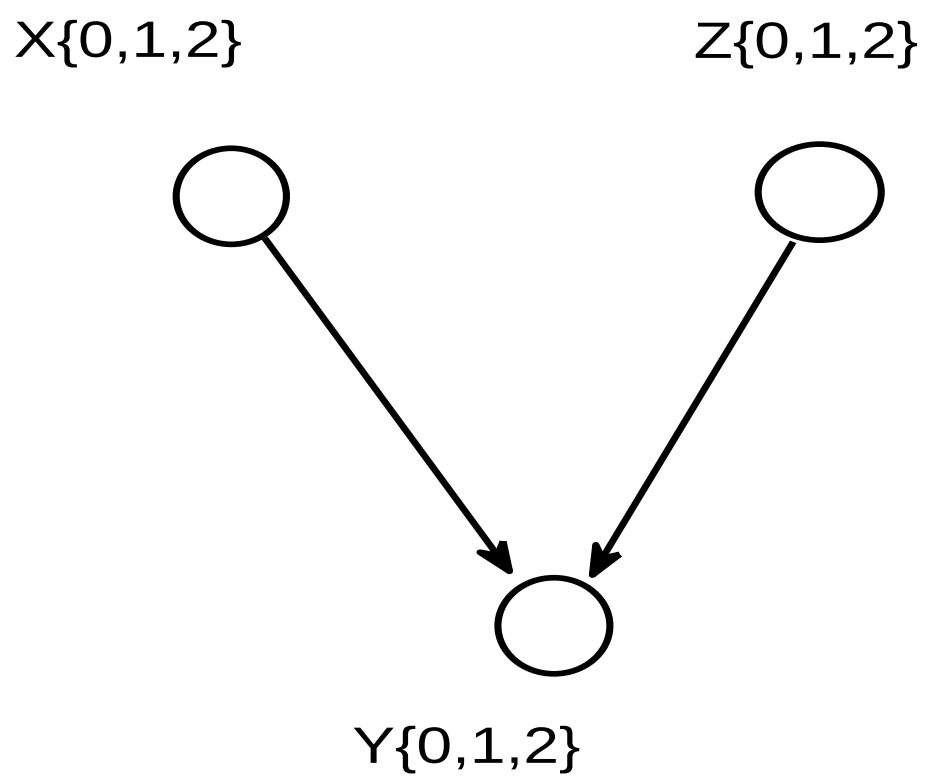

**Figure 1. DAG of 3-value X, 3-value Y, 3-value Z,** $X \perp Z$ **.**

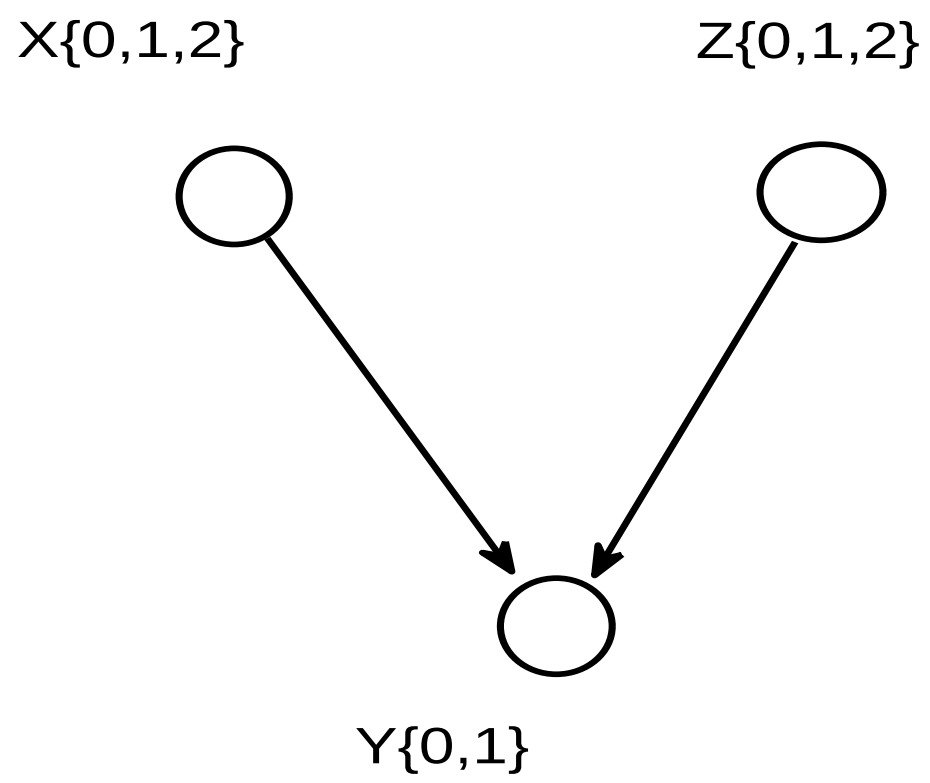

**Figure 2. DAG of 3-value X,2-value Y, 3-value Z,** $X \perp Z$ **.**

This causal model can describe the following problem: Let $X$ represent the degree of smoking, while " $X = 0$ " denote "never smoking", " $X = 1$ " denote "occasional smoking", " $X = 2$ " denote " frequently smoking"; Let $Y$ denote the different Stages of lung cancer, " $Y = 0$ " represent "no cancer", " $Y = 1$ " represent the "early stage of cancer", " $Y = 2$ " represent "later stage of cancer"; Let $Z$ denote the age of person, " $Z = 0$ " represent "adolescent", " $Z = 1$ " represent "middle-aged", " $Z = 2$ " represent "old". Similar to the discussion of [4], we have,

$$P_0(X=i)=P(X=i),\qquad P_0(Z=j)=P(Z=j),\quad i,j\in\{0,1,2\}$$
$$P_0(Y=k\mid X=0,Z=j)=P(Y=k\mid X=0,Z=j),\quad j,k\in\{0,1,2\}$$

Denote

$$a_1=P(X=1),\quad a_2=P(X=2),\quad P(X=0)=1-a_1-a_2=a_0;$$
$$c_1=P(Z=1),\quad c_2=P(Z=2),\quad P(Z=0)=1-c_1-c_2=c_0;$$
$$b_{ij}^k=P(Y=k\mid X=i,Z=j),\quad u_{ij}^k=P_0(Y=k\mid X=i,Z=j),\quad i,j,k=0,1,2$$

where, $\sum_{k=0}^{2} b_{ij}^k=1;\ \ \sum_{k=0}^{2} u_{ij}^k=1; u_{0j}^k=b_{0j}^k,\ j,k\in\{0,1,2\}$.

The identifiability of $P_0(Y)$ is to express $P_0(Y=k), k=0,1,2$ using $\{\{a_i\}_{i=0}^2, \{c_i\}_{i=0}^2, \{b_{0j}^k, j,k=0,1,2\}\}$.

As $P_0(Y=0)$ is identifiable, and the identifiability discussion of $P_0(Y=k), k=1,2$ is similar to $P_0(Y=0)$ , what we do is just to discuss the following intervention probability,

$$\begin{aligned}P_0(Y=1)&=\sum_{i=0}^{2}\sum_{j=0}^{2}P_0(Y=1\mid X=i,Z=j)P_0(X=i)P_0(Z=j)\\&=c_0(b_{00}^1a_0+u_{10}^1a_1+u_{20}^1a_2)+c_1(b_{01}^1a_0+u_{11}^1a_1+u_{21}^1a_2)+c_2(b_{02}^1a_0+u_{12}^1a_1+u_{22}^1a_2)\end{aligned}\tag{3.1}$$

Considering the following instrumental information:

(1) $X\perp Y:\ b_{00}^1c_0+b_{01}^1c_1+b_{02}^1c_2=u_{10}^1c_0+u_{11}^1c_1+u_{12}^1c_2=u_{20}^1c_0+u_{21}^1c_1+u_{22}^1c_2$

(2) $X\perp Y\mid Z=0\ \Rightarrow\ u_{10}^1=u_{20}^1=b_{00}^1$

(3) $X\perp Y\mid Z=1\ \Rightarrow\ u_{11}^1=u_{21}^1=b_{01}^1$

(4) $X\perp Y\mid Z=2\ \Rightarrow\ u_{12}^1=u_{22}^1=b_{02}^1$

(5) $Y\perp Z\ \Rightarrow\ b_{00}^1a_0+u_{10}^1a_1+u_{20}^1a_2=b_{01}^1a_0+u_{11}^1a_1+u_{21}^1a_2=b_{02}^1a_0+u_{12}^1a_1+u_{22}^1a_2$

(6) $Y\perp Z\mid X=0\ \Rightarrow\ b_{00}^1=b_{01}^1=b_{02}^1$

(7) $Y\perp Z\mid X=1\ \Rightarrow\ u_{10}^1=u_{11}^1=u_{12}^1$

(8) $Y\perp Z\mid X=2\ \Rightarrow\ u_{20}^1=u_{21}^1=u_{22}^1$

(9) $X\perp Z\mid Y=0\ \Rightarrow\ \dfrac{b_{00}^0}{u_{10}^0}=\dfrac{b_{01}^0}{u_{11}^0}=\dfrac{b_{02}^0}{u_{12}^0},\quad \dfrac{u_{10}^0}{u_{20}^0}=\dfrac{u_{11}^0}{u_{21}^0}=\dfrac{u_{12}^0}{u_{22}^0}$

(10) $X\perp Z\mid Y=1\ \Rightarrow\ \dfrac{b_{00}^1}{u_{10}^1}=\dfrac{b_{01}^1}{u_{11}^1}=\dfrac{b_{02}^1}{u_{12}^1},\quad \dfrac{u_{10}^1}{u_{20}^1}=\dfrac{u_{11}^1}{u_{21}^1}=\dfrac{u_{12}^1}{u_{22}^1}$

(11) $X\perp Z\mid Y=2\ \Rightarrow\ \dfrac{b_{00}^2}{u_{10}^2}=\dfrac{b_{01}^2}{u_{11}^2}=\dfrac{b_{02}^2}{u_{12}^2},\quad \dfrac{u_{10}^2}{u_{20}^2}=\dfrac{u_{11}^2}{u_{21}^2}=\dfrac{u_{12}^2}{u_{22}^2}$

### 3.1.2. Identifiability of Causal Model

According to the analysis of above instrumental information, we can obtain the following theorem:

**Theorem 5:** Suppose that $P_0$ satisfies one of the following conditions :

(a) $X\perp Y$;

OR (b) $X\perp Y\mid Z$;

OR (c) $Y \perp Z \bigcap X \perp Y \mid Z = i$, $i = 0,1,2$;

OR (d) $(X \perp Z \mid Y = 1) \bigcap (X \perp Y \mid Z = i)$, $i = 0,1,2$.

Then, $P_0(Y = 1)$ is identifiable, and,

$$P_0(Y=1)=\begin{cases} b_{00}^1 c_0 + b_{01}^1 c_1 + b_{02}^1 c_2, & X \perp Y \\ b_{00}^1 c_0 + b_{01}^1 c_1 + b_{02}^1 c_2, & X \perp Y \mid Z \\ b_{0i}^1, & Y \perp Z \bigcap X \perp Y \mid Z = i, \quad i = 0,1,2 \\ b_{00}^1, & Y \perp Z \mid X = 1 \bigcap X \perp Y \mid Z \\ b_{0i}^1, & (X \perp Z \mid Y = 1) \bigcap (X \perp Y \mid Z = i), \quad i = 0,1,2 \\ (\sum_{j=0}^{2} b_{0j}^1 c_j) a_0 + b_{0i}^1 (1 - a_0), & (Y \perp Z \mid X = 1 \bigcap Y \perp Z \mid X = 2) \bigcap (X \perp Y \mid Z = i), \\ & i = 0,1,2 \end{cases}$$

**Proof:**

a) Employing condition (1) $X \perp Y$ into (3.1), we can directly calculate and obtain the following result:

$$P_0(Y=1) = a_0(b_{00}^1 c_0 + b_{01}^1 c_1 + b_{02}^1 c_2) + a_1(b_{00}^1 c_0 + b_{01}^1 c_1 + b_{02}^1 c_2) + a_2(b_{00}^1 c_0 + b_{01}^1 c_1 + b_{02}^1 c_2)$$
$$= b_{00}^1 c_0 + b_{01}^1 c_1 + b_{02}^1 c_2$$

That is, $X \perp Y$ is the identifiability information of $P_0(Y = 1)$.

b) while conditions (2) $X \perp Y \mid Z = 0$, (3) $X \perp Y \mid Z = 1$, and (4) $X \perp Y \mid Z = 2$ hold, we can derive that $X \perp Y \mid Z$ holds, and then (1) $X \perp Y$ holds.

Then, we obtain

$$P_0(Y=1) = b_{00}^1 c_0 + b_{01}^1 c_1 + b_{02}^1 c_2,$$

Hence, $X \perp Y \mid Z$ is the identifiability instrumental information of $P_0(Y = 1)$.

c) While conditions (5) and (2) hold, we can obtain $Y \perp Z \bigcap X \perp Y \mid Z = 0$. Utilizing the conditions of parameters satisfying (5) and (2) to formula (3.1), we can obtain that,

$$P_0(Y=1) = b_{00}^1.$$

Similarly, Utilizing the conditions of parameters satisfying (5)(3) OR (5)(4), formula (3.1), we can obtain that,

$$P_0(Y=1) = b_{01}^1 \text{ OR } b_{02}^1.$$

Obviously, when (5) and (2)(3)(4) hold, $P_0(Y = 1) = b_{00}^1$.

d)Applying the parameter conditions of (10) and (2), OR (10) and (3), OR (10) and (4) to formula (3.1), we can obtain,

$$P_0(Y=1) = b_{00}^1 \text{ OR } b_{01}^1 \text{ OR } b_{02}^1.$$

Obviously, when (10) and (2)(3)(4) hold,

$$P_0(Y=1) = b_{00}^1.$$

e) When conditions (7)(8)(2) hold, applying the conditions of parameters to formula (3.1), we can obtain,

$$P_0(Y=1)=(\sum_{j=0}^{2} b_{0j}^1 c_j)a_0 + b_{00}^1(1-a_0)$$

Similarly, utilizing the parameter conditions of (7)(8)(3), OR (7)(8)(4) to formula (3.1), we can obtain ,

$$P_0(Y=1)=(\sum_{j=0}^{2} b_{0j}^1 c_j)a_0 + b_{01}^1(1-a_0) \text{ OR } (\sum_{j=0}^{2} b_{0j}^1 c_j)a_0 + b_{02}^1(1-a_0) .$$

□

## 3.2. " $X \perp Z$ ", Counterfactual causal graph with 3-value X, 2-value Y, 3-value Z

### 3.2.1. DAG model description and identifiable instrumental information

In this section, we discuss the counterfactual causal graphic model with 3-value X, 2-value Y, 3-value Z (Fig. 2), with sample spaces $X = Z = \{0,1,2\}, Y = \{0,1\}$ .

This causal model can describe the following problem: Let $X$ represent the degree of smoking, where " $X=0$ " represents " never smoking", " $X=1$ " represents "occasional smoking", and " $X=2$ " represents " frequently smoking"; Let $Y$ represent stages of cancer, where " $Y=0$ " represents "no lung cancer", " $Y=1$ " represents "lung cancer"; Let $Z$ represent ages, where " $Z=0$ " represents "adolescent", " $Z=1$ " represents "middle-aged", and " $Z=2$ " represents "old".

Denote

$$P_0(X=0)=a_0,\quad P_0(X=1)=a_1,\quad P_0(X=2)=a_2,$$
$$P_0(Z=0)=c_0,\quad P_0(Z=1)=c_1,\quad P_0(Z=2)=c_2$$
$$b_{ij}^k=P(Y=k|X=i,Z=j), u_{ij}^k=P_0(Y=k|X=i,Z=j), k=0,1; i,j=0,1,2.$$

where, $\sum_{k=0}^{1} b_{ij}^k = 1;\quad \sum_{k=0}^{1} u_{ij}^k = 1; u_{0j}^k = b_{0j}^k, i,j \in \{0,1,2\}, k \in \{0,1\}.$

The identifiability of $P_0(Y)$ is to express $P_0(Y=k), k=0,1$ using $\{\{a_i\}_{i=0}^2, \{c_i\}_{i=0}^2,$ $\{b_{0j}^k, j=0,1,2, k=0,1\}\}$ .

As $P_0(Y=0)$ is identifiable, we only discuss the identifiability of $P_0(Y=1)$, that is to discuss the following intervention probability,

$$\begin{aligned} P_0(Y=1) &= \sum_{i=0}^{2}\sum_{j=0}^{2} P_0(Y=1\,|\,X=i,Z=j)P_0(X=i)P_0(Z=j) \\ &= c_0(b_{00}^1 a_0 + u_{10}^1 a_1 + u_{20}^1 a_2) + c_1(b_{01}^1 a_0 + u_{11}^1 a_1 + u_{21}^1 a_2) + c_2(b_{02}^1 a_0 + u_{12}^1 a_1 + u_{22}^1 a_2) \end{aligned} \tag{3.2}$$

Considering the following instrumental information:

(1) $X \perp Y \Rightarrow\ b_{00}^1 c_0 + b_{01}^1 c_1 + b_{02}^1 c_2 = u_{10}^1 c_0 + u_{11}^1 c_1 + u_{12}^1 c_2 = u_{20}^1 c_0 + u_{21}^1 c_1 + u_{22}^1 c_2$

(2) $X \perp Y \,|\, Z=0 \ \Rightarrow u_{10}^1 = u_{20}^1 = b_{00}^1$

(3) $X \perp Y \,|\, Z=1 \ \Rightarrow u_{11}^1 = u_{21}^1 = b_{01}^1$

(4) $X \perp Y \,|\, Z=2 \ \Rightarrow u_{12}^1 = u_{22}^1 = b_{02}^1$

(5) $Y \perp Z \ \Rightarrow b_{00}^1 a_0 + u_{10}^1 a_1 + u_{20}^1 a_2 = b_{01}^1 a_0 + u_{11}^1 a_1 + u_{21}^1 a_2 = b_{02}^1 a_0 + u_{12}^1 a_1 + u_{22}^1 a_2$

$(6)\, Y \perp Z \mid X = 0 \ \Rightarrow b_{00}^1 = b_{01}^1 = b_{02}^1$

$(7)\, Y \perp Z \mid X = 1 \ \Rightarrow u_{10}^1 = u_{11}^1 = u_{12}^1$

$(8)\, Y \perp Z \mid X = 2 \ \Rightarrow u_{20}^1 = u_{21}^1 = u_{22}^1$

$(9)\, X \perp Z \mid Y = 0 \ \Rightarrow \frac{b_{00}^0}{u_{10}^0} = \frac{b_{01}^0}{u_{11}^0} = \frac{b_{02}^0}{u_{12}^0}, \quad \frac{u_{10}^0}{u_{20}^0} = \frac{u_{11}^0}{u_{21}^0} = \frac{u_{12}^0}{u_{22}^0}$

$(10)\, X \perp Z \mid Y = 1 \ \Rightarrow \frac{b_{00}^1}{u_{10}^1} = \frac{b_{01}^1}{u_{11}^1} = \frac{b_{02}^1}{u_{12}^1}, \quad \frac{u_{10}^1}{u_{20}^1} = \frac{u_{11}^1}{u_{21}^1} = \frac{u_{12}^1}{u_{22}^1}$

### 3.2.2. Identifiability of Causal Model

According to the analysis of above instrumental information, we can obtain the following theorem:

**Theorem 6**: Suppose that $P_0$ satisfies one of the following conditions :

(a) $X \perp Y$;

OR (b) $X \perp Y \mid Z$;

OR (c) $Y \perp Z \bigcap X \perp Y \mid Z = i, i = 0,1,2$ ;

OR (d) $(X \perp Z \mid Y = 1) \bigcap (X \perp Y \mid Z = i), i = 0,1,2$.

Then $P_0(Y = 1)$ is identifiable, and

$$P_0(Y=1) = \begin{cases} b_{00}^1 c_0 + b_{01}^1 c_1 + b_{02}^1 c_2, & X \perp Y \\ b_{00}^1 c_0 + b_{01}^1 c_1 + b_{02}^1 c_2, & X \perp Y \mid Z \\ b_{0i}^1, & Y \perp Z \bigcap X \perp Y \mid Z = i, \quad i = 0,1,2 \\ b_{00}^1, & Y \perp Z \mid X = 1 \bigcap X \perp Y \mid Z \\ b_{0i}^1, & (X \perp Z \mid Y = 1) \bigcap (X \perp Y \mid Z = i), \quad i = 0,1,2 \\ (\sum_{j=0}^{2} b_{0j}^1 c_j) a_0 + b_{0i}^1 (1 - a_0), & (Y \perp Z \mid X = 1 \bigcap Y \perp Z \mid X = 2) \bigcap (X \perp Y \mid Z = i), \\ & i = 0,1,2 \end{cases}$$

The proof is similar to Theorem 5 (omitted).

## 3.3 "$X \perp Z$", Counterfactual causal graph with 2-value X, 2-value Y, 3-value Z

### 3.3.1. DAG model description and identifiability instrumental information

In this section ,we address the causal graph model of three variables $X, Y, Z$ where $X, Y$ are 2-value variables , and $Z$ is a 3-value variable (Fig.3), $X = Y = \{0,1\}$, $Z = \{0,1,2\}$. Let the range spaces of $X$ and $Y$ are 0,1} (2-value), where $X = 0$ represents "no smoking", "$X = 1$" represents "smoking"; "$Y = 0$" represents "no lung cancer", and "$Y=1$" represents "lung cancer". The range space of $Z$ is $\{0,1,2\}$ (3-value), if it is endowed the meaning of ages, we can let "$Z=0$" represent " adolescent", "$Z=1$" represent "middle-aged", and "$Z=2$" represent "old".
Denote

$$a_0 = P(X=0), \quad a_1 = P(X=1);$$
$$c_1 = P(Z=1), \quad c_2 = P(Z=2), \quad P(Z=0) = 1 - c_1 - c_2 = c_0;$$
$$b_{ij}^k = P(Y=k \mid X=i, Z=j), \quad u_{ij}^k = P_0(Y=k \mid X=i, Z=j), \quad i,k=0,1; j=0,1,2.$$

Obviously, we have,

$$P_0(X=i) = P(X=i), \quad P_0(Z=j) = P(Z=j), \quad i \in \{0,1\}, j \in \{0,1,2\}$$
$$P_0(Y=k \mid X=0, Z=j) = P(Y=k \mid X=0, Z=j), \quad j \in \{0,1,2\}, k \in \{0,1\}$$

The identifiability of $P_0(Y)$ is to express $P_0(Y=k), k=0,1$ with $\{\{a_i\}_{i=0}^1, \{c_i\}_{i=0}^2,$ $\{b_{0j}^k, j=0,1,2, k=0,1\}\}$.

Since $P_0(Y=0)$ is identifiable, we only address the following intervention probability to calculate the identifiability of $P_0(Y=1)$.

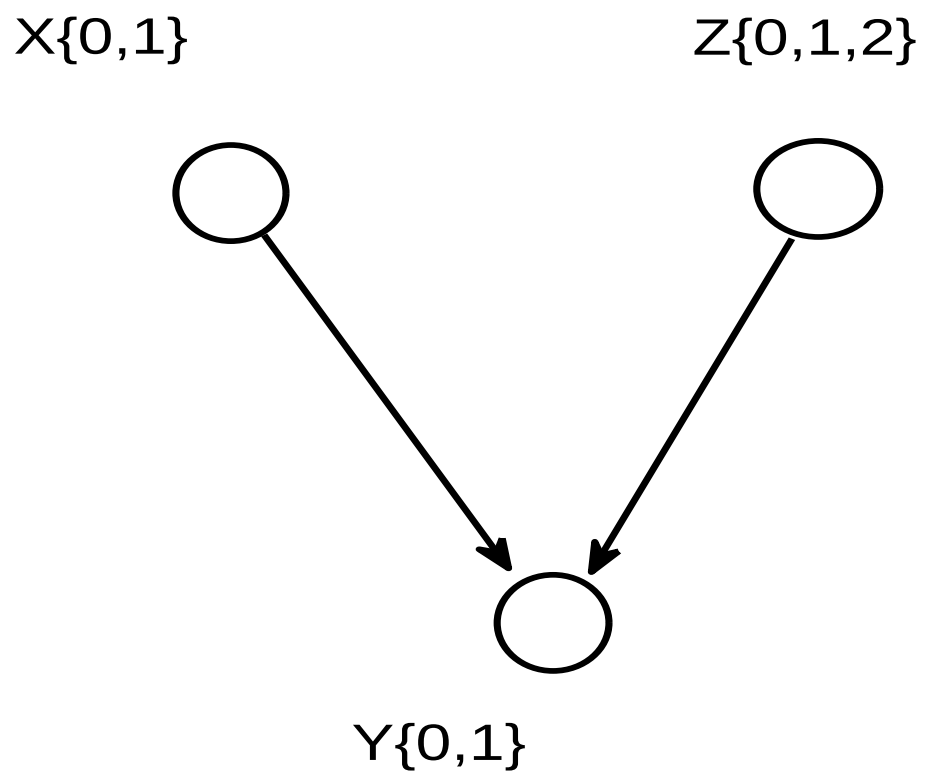

**Figure 3. Causal graph with 2-value X, 2-value Y, 3-value Z, and** $X \perp Z$ **.**

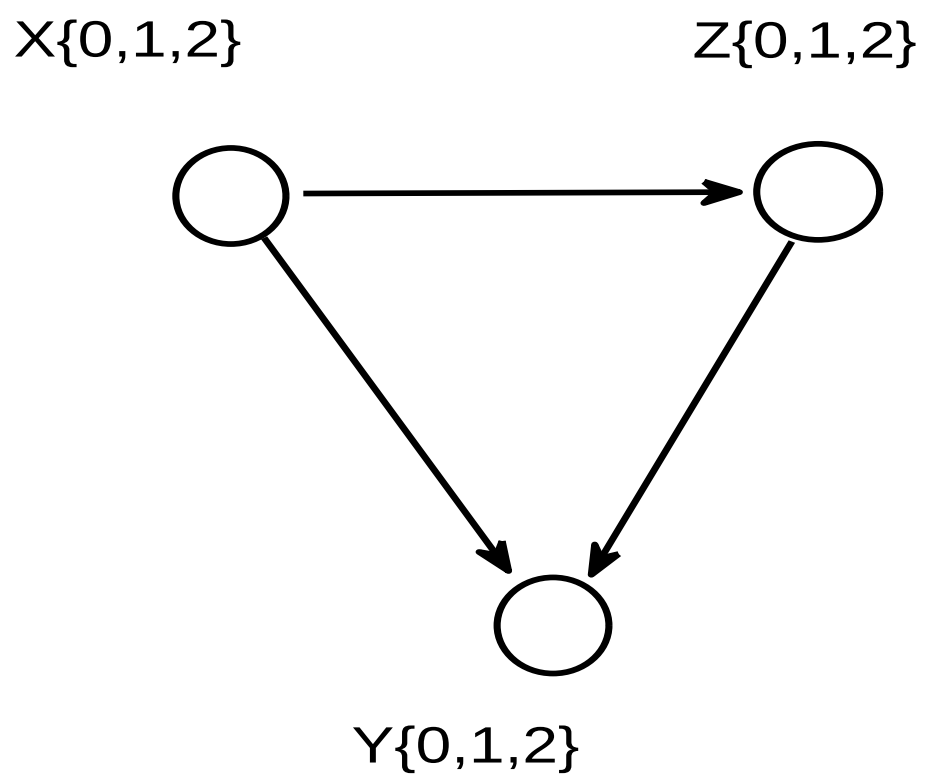

**Figure 4. Causal graph with 3-value X, 3-value Y, 3-value Z, and** $X \not\perp Z$ **.**

As,

$$\begin{aligned} P_0(Y=1) &= \sum_{i=0}^{1} \sum_{j=0}^{2} P_0(Y=1 \mid X=i, Z=j) P_0(X=i) P_0(Z=j) \\ &= c_0(b_{00}^1 a_0 + u_{10}^1 a_1) + c_1(b_{01}^1 a_0 + u_{11}^1 a_1) + c_2(b_{02}^1 a_0 + u_{12}^1 a_1) \end{aligned} \tag{3.3}$$

Considering the following instrumental information:

(1) $X \perp Y \Rightarrow \; b_{00}^1 c_0 + b_{01}^1 c_1 + b_{02}^1 c_2 = u_{10}^1 c_0 + u_{11}^1 c_1 + u_{12}^1 c_2$

(2) $X \perp Y \mid Z = 0 \;\; \Rightarrow \;\; u_{10}^1 = b_{00}^1$

(3) $X \perp Y \mid Z = 1 \;\; \Rightarrow \;\; u_{11}^1 = b_{01}^1$

(4) $X \perp Y \mid Z = 2 \;\; \Rightarrow \;\; u_{12}^1 = b_{02}^1$

(5) $Y \perp Z \;\; \Rightarrow \;\; b_{00}^1 a_0 + u_{10}^1 a_1 = b_{01}^1 a_0 + u_{11}^1 a_1 = b_{02}^1 a_0 + u_{12}^1 a_1$

(6) $Y \perp Z \mid X = 0 \;\; \Rightarrow \;\; b_{00}^1 = b_{01}^1 = b_{02}^1$

(7) $Y \perp Z \mid X = 1 \;\; \Rightarrow u_{10}^1 = u_{11}^1 = u_{12}^1$

(8) $X \perp Z \mid Y = 0 \;\; \Rightarrow \dfrac{b_{00}^0}{u_{10}^0} = \dfrac{b_{01}^0}{u_{11}^0} = \dfrac{b_{02}^0}{u_{12}^0}$

(9) $X \perp Z \mid Y = 1 \;\; \Rightarrow \dfrac{b_{00}^1}{u_{10}^1} = \dfrac{b_{01}^1}{u_{11}^1} = \dfrac{b_{02}^1}{u_{12}^1}$

### 3.3.2. Identifiability of Causal Model

According to the analysis of above instrumental information, we can obtain the following theorem:

**Theorem 7**: Suppose that $P_0$ satisfies one of the following conditions :

(a) $X \perp Y$ ;

OR (b) $X \perp Y \mid Z$ ;

OR (c) $Y \perp Z \bigcap X \perp Y \mid Z = i, i = 0,1,2$ ;

OR (d) $Y \perp Z \mid X = 1 \bigcap X \perp Y \mid Z = i, i = 0,1,2$ .

Then, $P_0(Y = 1)$ is identifiable, and

$$P_0(Y=1) = \begin{cases} b_{00}^1 c_0 + b_{01}^1 c_1 + b_{02}^1 c_2, & X \perp Y \\ b_{00}^1 c_0 + b_{01}^1 c_1 + b_{02}^1 c_2, & X \perp Y \mid Z \\ b_{0i}^1, & Y \perp Z \bigcap X \perp Y \mid Z = i, \quad i = 0,1,2 \\ b_{00}^1, & Y \perp Z \mid X = 1 \bigcap X \perp Y \mid Z \\ b_{0i}^1, & (X \perp Z \mid Y = 1) \bigcap (X \perp Y \mid Z = i), \quad i = 0,1,2 \\ (\sum_{j=0}^{2} b_{0j}^1 c_j) a_0 + b_{0i}^1 (1 - a_0), & (Y \perp Z \mid X = 1) \bigcap (X \perp Y \mid Z = i), \quad i = 0,1,2 \end{cases}$$

The proof is similar to Theorem 5 (omitted).

# 4. "$X \perp Z$", Extension of counterfactual causal graphic model with {X,Y,Z}

## 4.1. $X \perp Z$ ,Causal graph with 3-value X, 3-value Y, 3-value Z

### 4.1.1. DAG model description and identifiability instrumental information

For the counterfactual causal graphic model with 3-value X, 3-value Y, 3-value Z, and $X \perp Z$ (Fig.4), the sample spaces are $X = Z = Y = \{0,1,2\}$. The background of this causal model can be treated as the simple extension of reference [5]. Let $X$ represent the three dosage levels of soil fertilizer, where "X=0", "X=1" and "X=2" represent the dosage from less to more. For example, "X=0" denotes "no fertilization", "X=1" denotes "little level" , and "X=2" denotes "proper level". (Obviously, "X" can be expressed by the real dosage of fertilization, for example, 0, 50 and 100 kilograms of a hectare). Let $Y$ represent the outcome of beans with different dosage of fertilization, where "Y=0", "Y=1", and "Y=2" represent the "low-yield", "medium-yield" and "high-yield " respectively. Let "Z" represent the effect of fertilizer to a kind of microbe's amount in the soil so that it can affect the yield of bean, where "Z=0","Z=1", and "Z=2" represent the "little", "normal" and "much" levels of microbe respectively.

Denote

$$a_0 = P(X=0), \quad a_1 = P(X=1), \quad P(X=2) = a_2 = 1 - a_0 - a_1;$$
$$b_{ij}^k = P(Y=k \mid X=i, Z=j), \quad u_{ij}^k = P_0(Y=k \mid X=i, Z=j),$$
$$c_{ij} = P_0(Z=j \mid X=i), \quad i,j,k = 0,1,2.$$

And,

$$P_0(X=i) = P(X=i), \quad P_0(Z=j \mid X=0) = P(Z=j \mid X=0),$$
$$P_0(Y=k \mid X=0, Z=j) = P(Y=k \mid X=0, Z=j), \quad i,j,k \in \{0,1,2\}$$

The identifiability of $P_0(Y)$ is to express $P_0(Y=k), k=0,1,2$ using $\{\{a_i\}_{i=0}^2, \{c_{0j}\}_{j=0}^2, \{b_{0j}^k, j,k=0,1,2\}\}$. Since $P_0(Y=0)$ is identifiable, the identifiability discussion of $P_0(Y=k), k=1,2$ is only relative to the following intervention probability,

$$\begin{aligned} P_0(Y=1) &= \sum_{i=0}^2 \sum_{j=0}^2 P_0(Y=1 \mid X=i, Z=j) P_0(Z=j \mid X=i) P_0(X=i) \\ &= (\sum_{j=0}^2 b_{0j}^1 c_{0j}) a_0 + (\sum_{j=0}^2 u_{1j}^1 c_{1j}) a_1 + (\sum_{j=0}^2 u_{2j}^1 c_{2j}) a_2 \end{aligned} \tag{4.1}$$

Considering the following instrumental information:

(1) $X \perp Y \Rightarrow b_{00}^1 c_{00} + b_{01}^1 c_{01} + b_{02}^1 c_{02} = u_{10}^1 c_{10} + u_{11}^1 c_{11} + u_{12}^1 c_{12} = u_{20}^1 c_{20} + u_{21}^1 c_{21} + u_{22}^1 c_{22}$

(2) $X \perp Y \mid Z=0 \;\; \Rightarrow b_{00}^1 = u_{10}^1 = u_{20}^1$

(3) $X \perp Y \mid Z=1 \;\; \Rightarrow b_{01}^1 = u_{11}^1 = u_{21}^1$

(4) $X \perp Y \mid Z=2 \;\; \Rightarrow b_{02}^1 = u_{12}^1 = u_{22}^1$

(5) $Y \perp Z \Rightarrow$

$$\frac{b_{00}^1 c_{00} a_0 + u_{10}^1 c_{10} a_1 + u_{20}^1 c_{20} a_2}{c_{00} a_0 + c_{10} a_1 + c_{20} a_2} = \frac{b_{01}^1 c_{01} a_0 + u_{11}^1 c_{11} a_1 + u_{21}^1 c_{21} a_2}{c_{01} a_0 + c_{11} a_1 + c_{21} a_2} = \frac{b_{02}^1 c_{02} a_0 + u_{12}^1 c_{12} a_1 + u_{22}^1 c_{22} a_2}{c_{02} a_0 + c_{12} a_1 + c_{22} a_2}$$

(6) $Y \perp Z \mid X = 0 \Rightarrow b_{00}^1 = b_{01}^1 = b_{02}^1$

(7) $Y \perp Z \mid X = 1 \Rightarrow u_{10}^1 = u_{11}^1 = u_{12}^1$

(8) $Y \perp Z \mid X = 2 \Rightarrow u_{20}^1 = u_{21}^1 = u_{22}^1$

(9) $X \perp Z \Rightarrow c_{00} = c_{10} = c_{20}, \quad c_{01} = c_{11} = c_{21}, \quad c_{02} = c_{12} = c_{22}$

(10) $X \perp Z \mid Y = 0 \Rightarrow \dfrac{b_{00}^0 c_{00}}{u_{10}^0 c_{10}} = \dfrac{b_{01}^0 c_{01}}{u_{11}^0 c_{11}} = \dfrac{b_{02}^0 c_{02}}{u_{12}^0 c_{12}}, \quad \dfrac{u_{10}^0 c_{10}}{u_{20}^0 c_{20}} = \dfrac{u_{11}^0 c_{11}}{u_{21}^0 c_{21}} = \dfrac{u_{12}^0 c_{12}}{u_{22}^0 c_{22}}$

(11) $X \perp Z \mid Y = 1 \Rightarrow \dfrac{b_{00}^1 c_{00}}{u_{10}^1 c_{10}} = \dfrac{b_{01}^1 c_{01}}{u_{11}^1 c_{11}} = \dfrac{b_{02}^1 c_{02}}{u_{12}^1 c_{12}}, \quad \dfrac{u_{10}^1 c_{10}}{u_{20}^1 c_{20}} = \dfrac{u_{11}^1 c_{11}}{u_{21}^1 c_{21}} = \dfrac{u_{12}^1 c_{12}}{u_{22}^1 c_{22}}$

(12) $X \perp Z \mid Y = 2 \Rightarrow \dfrac{b_{00}^2 c_{00}}{u_{10}^2 c_{10}} = \dfrac{b_{01}^2 c_{01}}{u_{11}^2 c_{11}} = \dfrac{b_{02}^2 c_{02}}{u_{12}^2 c_{12}}, \quad \dfrac{u_{10}^2 c_{10}}{u_{20}^2 c_{20}} = \dfrac{u_{11}^2 c_{11}}{u_{21}^2 c_{21}} = \dfrac{u_{12}^2 c_{12}}{u_{22}^2 c_{22}}$

### 4.1.2. Identifiability of Causal Model

According to the analysis of above instrumental information, we can obtain the following theorem:

**Theorem 8**: Suppose that $P_0$ satisfies one of the following conditions:

(a) $X \perp Y$ ;

OR (b) $X \perp Y \mid Z \bigcap Y \perp Z \mid X = i, \quad i = 0,1,2$ ;

OR (c) $(Y \perp Z \mid X = 1 \bigcap Y \perp Z \mid X = 2) \bigcap (X \perp Y \mid Z = j), \quad j = 0,1,2$.

Then, $P_0(Y = 1)$ is identifiable, and

$$P_0(Y=1) = \begin{cases} b_{00}^1 c_{00} + b_{01}^1 c_{01} + b_{02}^1 c_{02}, & X \perp Y \\ b_{00}^1, & X \perp Y \mid Z \bigcap (Y \perp Z \mid X = i), \quad i = 0,1,2 \\ (\sum_{j=0}^{2} b_{0j}^1 c_{0j}) a_0 + b_{0i}^1 (1 - a_0), & (Y \perp Z \mid X = 1 \bigcap Y \perp Z \mid X = 2) \bigcap (X \perp Y \mid Z = i), \\ & i = 0,1,2 \end{cases}$$

**Proof**:

a) Applying the parameter condition satisfying (1) $X \perp Y$ to formula (4.1), we can calculate that:

$$P_0(Y=1) = (\sum_{j=0}^{2} b_{0j}^1 c_{0j}) a_0 + (\sum_{j=0}^{2} u_{1j}^1 c_{1j}) a_1 + (\sum_{j=0}^{2} u_{2j}^1 c_{2j}) a_2 = b_{00}^1 c_{00} + b_{01}^1 c_{01} + b_{02}^1 c_{02}$$

b) When conditions (2)(3)(4) and (6) hold, applying the conditions of parameter to formula (4.1), we obtain

$$P_0(Y=1) = b_{00}^1 .$$

Applying the conditions of parameter satisfying (2)(3)(4) and (7) to formula (4.1), we

obtain

$$P_0(Y=1)=b_{00}^1 .$$

Applying the conditions of parameter satisfying (2)(3)(4) and (8) to formula (4.1), we obtain

$$P_0(Y=1)=b_{00}^1 .$$

c) When conditions (7)(8) and (2) hold, we can directly calculate

$$P_0(Y=1)=(b_{00}^1c_{00}+b_{01}^1c_{01}+b_{02}^1c_{02})a_0+b_{00}^1(1-a_0) .$$

When conditions (7)(8) and (3) hold, we can directly calculate

$$P_0(Y=1)=(b_{00}^1c_{00}+b_{01}^1c_{01}+b_{02}^1c_{02})a_0+b_{01}^1(1-a_0) .$$

When conditions (7)(8) and (4) hold, we can directly calculate

$$P_0(Y=1)=(b_{00}^1c_{00}+b_{01}^1c_{01}+b_{02}^1c_{02})a_0+b_{02}^1(1-a_0) .$$

□

## 4.2. $X \perp Z$ ,Causal graph with 3-value X, 2-value Y, 3-value Z

### 4.2.1. DAG model description and identifiability instrumental information

For the counterfactual causal graphic model with 3-value $X$ , 3-value $Z$ , 2-value $Y$ , and $X \perp Z$ , the ranges of space are $X=Z=\{0,1,2\}, Y=\{0,1\}$ (Fig. 5). We can explain the causal model as follows: “X=0” represents “ no fertilization”, “X=1” represents “little fertilization” , and “X=2” represents “proper fertilization”; “Y=0” represents the “low-yield” of bean , “Y=1” represents “high-yield”; And, “Z=0”, “Z=1”, “Z=2” represent the amount of microbe in the soil, which are the “little”, “normal” and “much” levels of microbe respectively.

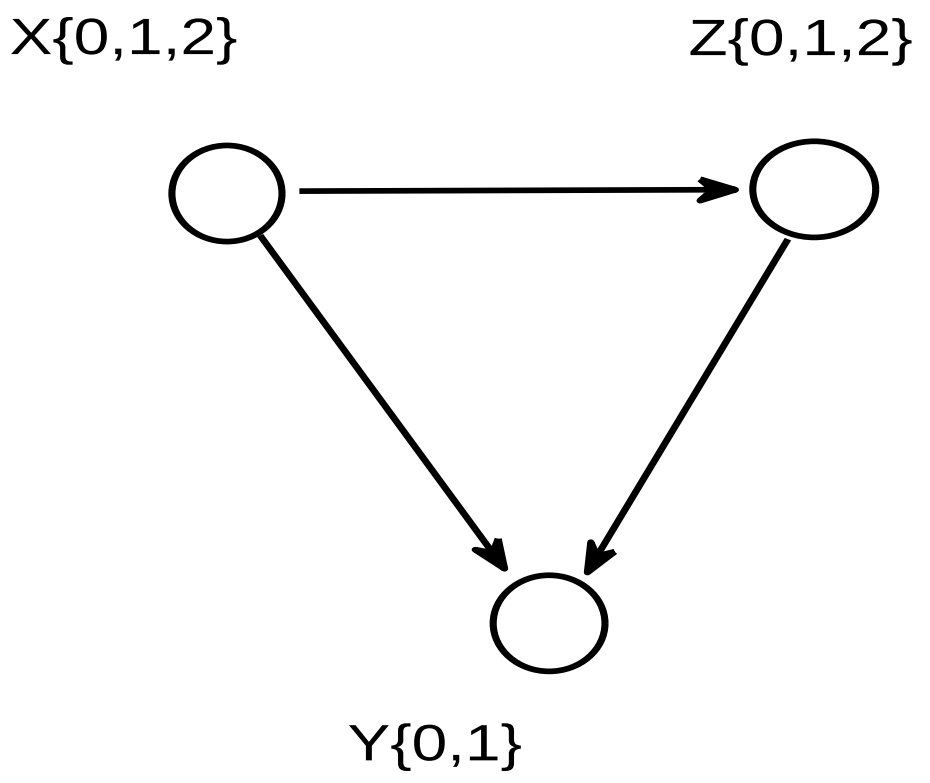

**Figure 5. Causal graph with 3-value X, 2-value Y, 3-value Z, and $X \perp Z$ .**

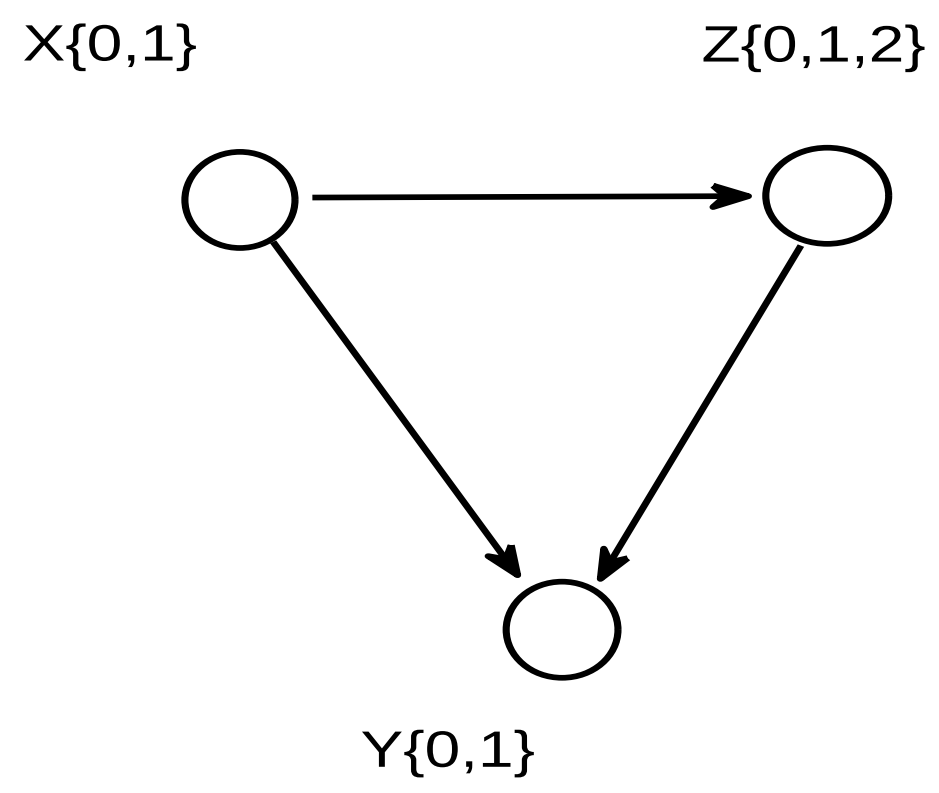

**Figure 6. Causal graph with 2-value X, 2-value Y, 3-value Z, and $X \perp Z$ .**

Denote,

$$a_0 = P(X=0),\quad a_1 = P(X=1),\quad a_2 = P(X=2) = 1 - a_0 - a_1;$$
$$b_{ij}^k = P(Y=k \mid X=i, Z=j),\quad u_{ij}^k = P_0(Y=k \mid X=i, Z=j),$$
$$c_{ij} = P_0(Z=j \mid X=i),\quad i,j = 0,1,2;\quad k=0,1.$$

where,, $b_{ij}^0 = 1 - b_{ij}^1$ , $u_{ij}^0 = 1 - u_{ij}^1$ .

Then,

$$P_0(X=i) = P(X=i),\qquad i = 0,1,2,$$
$$P_0(Y=k \mid X=0, Z=j) = P(Y=k \mid X=0, Z=j),\quad j=0,1,2, k=0,1.$$

To discuss the identifiability of $P_0(Y)$ is to express $P_0(Y=k), k=0,1$ using $\{\{a_i\}_{i=0}^2, \{c_{0j}\}_{j=0}^2,\ \{b_{0j}^k, j=0,1,2, k=0,1\}\}$.

As, $P_0(Y=0)$ is identifiable, the identifiability of $P_0(Y=1)$ is to consider the following intervention probability,

$$\begin{aligned}P_0(Y=1) &= \sum_{i=0}^{2}\sum_{j=0}^{2} P_0(Y=1 \mid X=i, Z=j) P_0(Z=j \mid X=i) P_0(X=i) \\ &= (b_{00}^1 c_{00} + b_{01}^1 c_{01} + b_{02}^1 c_{02}) a_0 + (u_{10}^1 c_{10} + u_{11}^1 c_{11} + u_{12}^1 c_{12}) a_1 + (u_{20}^1 c_{20} + u_{21}^1 c_{21} + u_{22}^1 c_{22}) a_2\end{aligned} \qquad (4.2)$$

Considering the following instrumental information:

(1) $X \perp Y \Rightarrow\ b_{00}^1 c_{00} + b_{01}^1 c_{01} + b_{02}^1 c_{02} = u_{10}^1 c_{10} + u_{11}^1 c_{11} + u_{12}^1 c_{12} = u_{20}^1 c_{20} + u_{21}^1 c_{21} + u_{22}^1 c_{22}$

(2) $X \perp Y \mid Z=0\ \Rightarrow u_{10}^1 = b_{00}^1 = u_{20}^1$

(3) $X \perp Y \mid Z=1\ \Rightarrow u_{11}^1 = u_{21}^1 = b_{01}^1$

(4) $X \perp Y \mid Z=2\ \Rightarrow u_{12}^1 = u_{22}^1 = b_{02}^1$

(5) $Y \perp Z\ \Rightarrow$

$$\frac{b_{00}^1 c_{00} a_0 + u_{10}^1 c_{10} a_1 + u_{20}^1 c_{20} a_2}{c_{00} a_0 + c_{10} a_1 + c_{20} a_2} = \frac{b_{01}^1 c_{01} a_0 + u_{11}^1 c_{11} a_1 + u_{21}^1 c_{21} a_2}{c_{01} a_0 + c_{11} a_1 + c_{21} a_2} = \frac{b_{02}^1 c_{02} a_0 + u_{12}^1 c_{12} a_1 + u_{22}^1 c_{22} a_2}{c_{02} a_0 + c_{12} a_1 + c_{22} a_2}$$

(6) $Y \perp Z \mid X=0\ \Rightarrow b_{00}^1 = b_{01}^1 = b_{02}^1$

(7) $Y \perp Z \mid X=1\ \Rightarrow u_{10}^1 = u_{11}^1 = u_{12}^1$

(8) $Y \perp Z \mid X = 2 \quad \Rightarrow \quad u_{20}^1 = u_{21}^1 = u_{22}^1$

(9) $X \perp Z \quad \Rightarrow c_{00} = c_{10} = c_{20}, \quad c_{01} = c_{11} = c_{21}, \quad c_{02} = c_{12} = c_{22}$

(10) $X \perp Z \mid Y = 0 \quad \Rightarrow \dfrac{b_{00}^0 c_{00}}{u_{10}^0 c_{10}} = \dfrac{b_{01}^0 c_{01}}{u_{11}^0 c_{11}} = \dfrac{b_{02}^0 c_{02}}{u_{12}^0 c_{12}}, \quad \dfrac{u_{10}^0 c_{10}}{u_{20}^0 c_{20}} = \dfrac{u_{11}^0 c_{11}}{u_{21}^0 c_{21}} = \dfrac{u_{12}^0 c_{12}}{u_{22}^0 c_{22}}$

(11) $X \perp Z \mid Y = 1 \quad \Rightarrow \dfrac{b_{00}^1 c_{00}}{u_{10}^1 c_{10}} = \dfrac{b_{01}^1 c_{01}}{u_{11}^1 c_{11}} = \dfrac{b_{02}^1 c_{02}}{u_{12}^1 c_{12}}, \quad \dfrac{u_{10}^1 c_{10}}{u_{20}^1 c_{20}} = \dfrac{u_{11}^1 c_{11}}{u_{21}^1 c_{21}} = \dfrac{u_{12}^1 c_{12}}{u_{22}^1 c_{22}}$

### 4.2.2. Identifiability of Causal Model

According to the analysis of above instrumental information, we can obtain the following theorem:

**Theorem 9**: Suppose that $P_0$ satisfies one of the following conditions:

(a) $X \perp Y$ ;

OR (b) $X \perp Y \mid Z \bigcap Y \perp Z \mid X = i, \quad i = 0,1,2$ ;

OR (c) $(Y \perp Z \mid X = 1 \bigcap Y \perp Z \mid X = 2) \bigcap (X \perp Y \mid Z = j), j = 0,1,2$.

Then, $P_0(Y=1)$ is identifiable, and

$$P_0(Y=1) = \begin{cases} b_{00}^1 c_{00} + b_{01}^1 c_{01} + b_{02}^1 c_{02}, & X \perp Y \\ b_{00}^1, & (X \perp Y \mid Z) \bigcap (Y \perp Z \mid X = i), \quad i = 0,1,2 \\ (\sum_{j=0}^{2} b_{0j}^1 c_{0j}) a_0 + b_{0i}^1 (1 - a_0), & (Y \perp Z \mid X = 1 \bigcap Y \perp Z \mid X = 2) \bigcap (X \perp Y \mid Z = i), \\ & i = 0,1,2 \end{cases}$$

The proof is similar to Theorem 8 (omitted).

## 4.3. "$X \perp Z$", Causal graph with 2-value X, 2-value Y,3-value Z

### *4.3.1. DAG model description and identifiability instrumental information*

For the counterfactual causal graphic model with 2-value X, 2-value Y, 3-value Z, and $X \perp Z$ (Fig. 6), their sample spaces are $X = Y = \{0,1\}, Z = \{0,1,2\}$. The background model can be explained that："X=0" represents "no fertilization", "X=1" represents "using fertilization"; "Y=0" represents "low-yield" of bean, , "Y=1" represents "high-yield" of bean; And , "Z=0","Z=1","Z=2" represent the effect of fertilizer to microbe in the soil, which are the "little", "normal" and "much" levels of microbe respectively.

Denote

$$a_0 = P(X=0), \quad a_1 = P(X=1); \quad c_{ij} = P_0(Z=j \mid X=i),$$

$$b_{ij}^k = P(Y=k \mid X=i, Z=j), \quad u_{ij}^k = P_0(Y=k \mid X=i, Z=j), \quad j = 0,1,2; \quad i,k = 0,1.$$

where, $b_{ij}^0 = 1 - b_{ij}^1$, $u_{ij}^0 = 1 - u_{ij}^1$. And,

$$P_0(X = i) = P(X = i), \qquad i \in \{0,1\}$$
$$P_0(Y = k \mid X = 0, Z = j) = P(Y = k \mid X = 0, Z = j), \quad j \in \{0,1,2\}, k \in \{0,1\}$$

Again, to investigate the identifiability of $P_0(Y)$ is to express $P_0(Y = k), k = 0,1$ using $\{\{a_i\}_{i=0}^1, \{c_{0j}\}_{j=0}^2, \{b_{0j}^k, j = 0,1,2, k = 0,1\}\}$. Since, $P_0(Y = 0)$ is identifiable, to discuss the identifiability of $P_0(Y = 1)$, we should discuss the following intervention probability,

$$\begin{aligned} P_0(Y = 1) &= \sum_{i=0}^{1}\sum_{j=0}^{2} P_0(Y = 1 \mid X = i, Z = j) P_0(Z = j \mid X = i) P_0(X = i) \\ &= (b_{00}^1 c_{00} + b_{01}^1 c_{01} + b_{02}^1 c_{02}) a_0 + (u_{10}^1 c_{10} + u_{11}^1 c_{11} + u_{12}^1 c_{12}) a_1 \end{aligned} \tag{4.3}$$

Considering the following instrumental information:

(1) $X \perp Y \Rightarrow b_{00}^1 c_{00} + b_{01}^1 c_{01} + b_{02}^1 c_{02} = u_{10}^1 c_{10} + u_{11}^1 c_{11} + u_{12}^1 c_{12}$

(2) $X \perp Y \mid Z = 0 \Rightarrow u_{10}^1 = b_{00}^1$

(3) $X \perp Y \mid Z = 1 \Rightarrow u_{11}^1 = b_{01}^1$

(4) $X \perp Y \mid Z = 2 \Rightarrow u_{12}^1 = b_{02}^1$

(5) $Y \perp Z \Rightarrow \dfrac{b_{00}^1 c_{00} a_0 + u_{10}^1 c_{10} a_1}{c_{00} a_0 + c_{10} a_1} = \dfrac{b_{01}^1 c_{01} a_0 + u_{11}^1 c_{11} a_1}{c_{01} a_0 + c_{11} a_1} = \dfrac{b_{02}^1 c_{02} a_0 + u_{12}^1 c_{12} a_1}{c_{02} a_0 + c_{12} a_1}$

(6) $Y \perp Z \mid X = 0 \Rightarrow b_{00}^1 = b_{01}^1 = b_{02}^1$

(7) $Y \perp Z \mid X = 1 \Rightarrow u_{10}^1 = u_{11}^1 = u_{12}^1$

(8) $X \perp Z \Rightarrow c_{00} = c_{10}, \quad c_{01} = c_{11}, \quad c_{02} = c_{12}$

(9) $X \perp Z \mid Y = 0 \Rightarrow \dfrac{b_{00}^0 c_{00}}{u_{10}^0 c_{10}} = \dfrac{b_{01}^0 c_{01}}{u_{11}^0 c_{11}} = \dfrac{b_{02}^0 c_{02}}{u_{12}^0 c_{12}}$

(11) $X \perp Z \mid Y = 1 \Rightarrow \dfrac{b_{00}^1 c_{00}}{u_{10}^1 c_{10}} = \dfrac{b_{01}^1 c_{01}}{u_{11}^1 c_{11}} = \dfrac{b_{02}^1 c_{02}}{u_{12}^1 c_{12}}$

## 4.3.2. Identifiability of Causal Model

According to the analysis of above instrumental information, we can obtain the following theorem:

**Theorem 10**: Suppose that $P_0$ satisfies one of the following conditions

(a) $X \perp Y$ ;

OR (b) $X \perp Y \mid Z \bigcap Y \perp Z \mid X = i, \quad i = 0,1$ ;

OR (c) $(Y \perp Z \mid X = 1) \bigcap (X \perp Y \mid Z = j), j = 0,1,2$,

Then, $P_0(Y = 1)$ is identifiable, and

$$P_0(Y=1)=\begin{cases} b_{00}^1 c_{00}+b_{01}^1 c_{01}+b_{02}^1 c_{02}, & X\perp Y \\ b_{00}^1, & (X\perp Y\mid Z)\bigcap(Y\perp Z\mid X=i),\quad i=0,1 \\ (\sum_{j=0}^{2} b_{0j}^1 c_{0j})a_0+b_{0i}^1(1-a_0), & (Y\perp Z\mid X=1)\bigcap(X\perp Y\mid Z=i),\quad i=0,1,2 \end{cases}$$

The proof is similar to Theorem 8 (omitted).

# 5. Conclusions

Extending the three variables counterfactual causal graph from 2-value to 3-value in two cases, independence or dependence of control variable and instrumental variable, are totally 14 cases. Limited to the paper length, we only give 6 cases of counterfactual causal graph model, and obtain the sufficient identifiability condition of $P_0(Y=1)$. Further research will focus on (1) the sufficient and necessary identifiability condition of $P_0(Y=1)$, and discuss sufficient and necessary identifiability conditions of $P_0(Y)$. (2) the identifiability condition of $P_0(Y)$ with the 3-variable $K$-value ($K>3$) counterfactual causal graphic model. (3) the identifiability conditions of counterfactual causal graphic model with multi-value intervention.

# 6. Acknowledgements

The author would like to thank Ms. Wei He, Ms. ShanShan Han and Prof. Zhongguo Zheng (Peking University) for the valuable discussions.